\documentclass[prl,twocolumn,superscriptaddress]{revtex4}
\usepackage{epsfig}
\usepackage{times}
\usepackage{amsmath}
\usepackage{amsfonts}
\usepackage{amssymb}
\usepackage[usenames]{color}
\usepackage{graphicx}

\newcommand{\beq}{\begin{equation}}
\newcommand{\eeq}{\end{equation}}
\newcommand{\beqa}{\begin{eqnarray}}
\newcommand{\eeqa}{\end{eqnarray}}

\begin{document}
\title{Entangling unstable optically active matter qubits}
\author{Yuichiro Matsuzaki}
\affiliation{Department of Materials, University of Oxford, OX1 3PH, U. K.}
\author{Simon C. Benjamin\footnote{s.benjamin@qubit.org}}
\affiliation{Department of Materials, University of Oxford, OX1 3PH, U. K.}
\affiliation{Centre for Quantum Technologies, National University of Singapore, 3 Science Drive 2, Singapore 117543.}
\author{Joseph Fitzsimons}
\affiliation{Department of Materials, University of Oxford, OX1 3PH, U. K.}

\begin{abstract}
In distributed quantum computation, small devices composed of a single or a few qubits
 are networker together to achieve a scalable machine.
Typically there is an optically active matter qubit at each node, so that photons are exploited to achieve remote entanglement. 
 However, in many systems the optically active states are unstable or
 poorly defined.  We report
a
 scheme to perform a high-fidelity entanglement
 operation even given severe instability. The protocol exploits the existence of optically excited states for phase acquisition without actually exciting those states; it functions with or without cavities and does not require number resolving detectors.
 
\end{abstract}

\maketitle

  One promising approach to quantum information processing (QIP) is
 {\em distributed} QIP~\cite{CEHM01a,BK01a,LBK01a,BPH01a,BKPV01a,Fengetal01a,Benjamin:2006p358}. Here scalability is achieved by networking many
  elementary nodes.
Qubits in remote nodes are coupled
 through an entanglement operation (EO) which typically utilizes some kind of photon interference effect. Most of the proposed EOs are implemented not
 deterministicly but rather are probabilistic in nature. Failure of the EO will be `heralded', i.e. the operator is aware of the failure, but in that case
 the qubits acted upon are effectively corrupt and need to be reset.
 Therefore if only one qubit is present at each node, performing an EO between
 two specific qubits implies a significant risk
of losing any prior entanglement with other qubits.  
 
There are many existing suggestions for implementing
 EOs, including a number of so-called path-erasure schemes~\cite{CEHM01a,BK01a,LBK01a,BPH01a,BKPV01a,BES01a,laddetal2006hybrid,PRAmatsuzaki2010distributed}.
 These approaches typically involve exciting an optical transition in the matter system at each node.
 However, in many real systems such transitions are inherently unstable due to energy
 fluctuation of the excited states. If such states exist in superposition with lower lying states, even briefly, then their instability will cause dephasing
and hence ultimately degradation of the entanglement operation~\cite{kaer2010non,reithmaieretal2004strong}.
 So one has to look for a robust way to generate high-fidelity
 entanglement under the effect of
 such energy
 fluctuations.
 
 Recently it has been shown that one can suppress such dephasing through temporal
 postselecton of emitted photons at the expense of decreasing success probability~\cite{nazir2009overcoming}. However,
 this scheme is sensitive to imperfections in the photodetector. 
 Moreover, due to
 the inherently low success probability,
 dark
 counts will be a relevant problem and will lead to decreased
 fidelity~\cite{CB01a}. Here we present a fully scalable procedure for distributed quantum computation
 by constructing a high-fidelity EO scheme which is relatively robust against such issues.
 
      \begin{figure}[h!]
   \begin{center}
    \includegraphics[width=7cm]{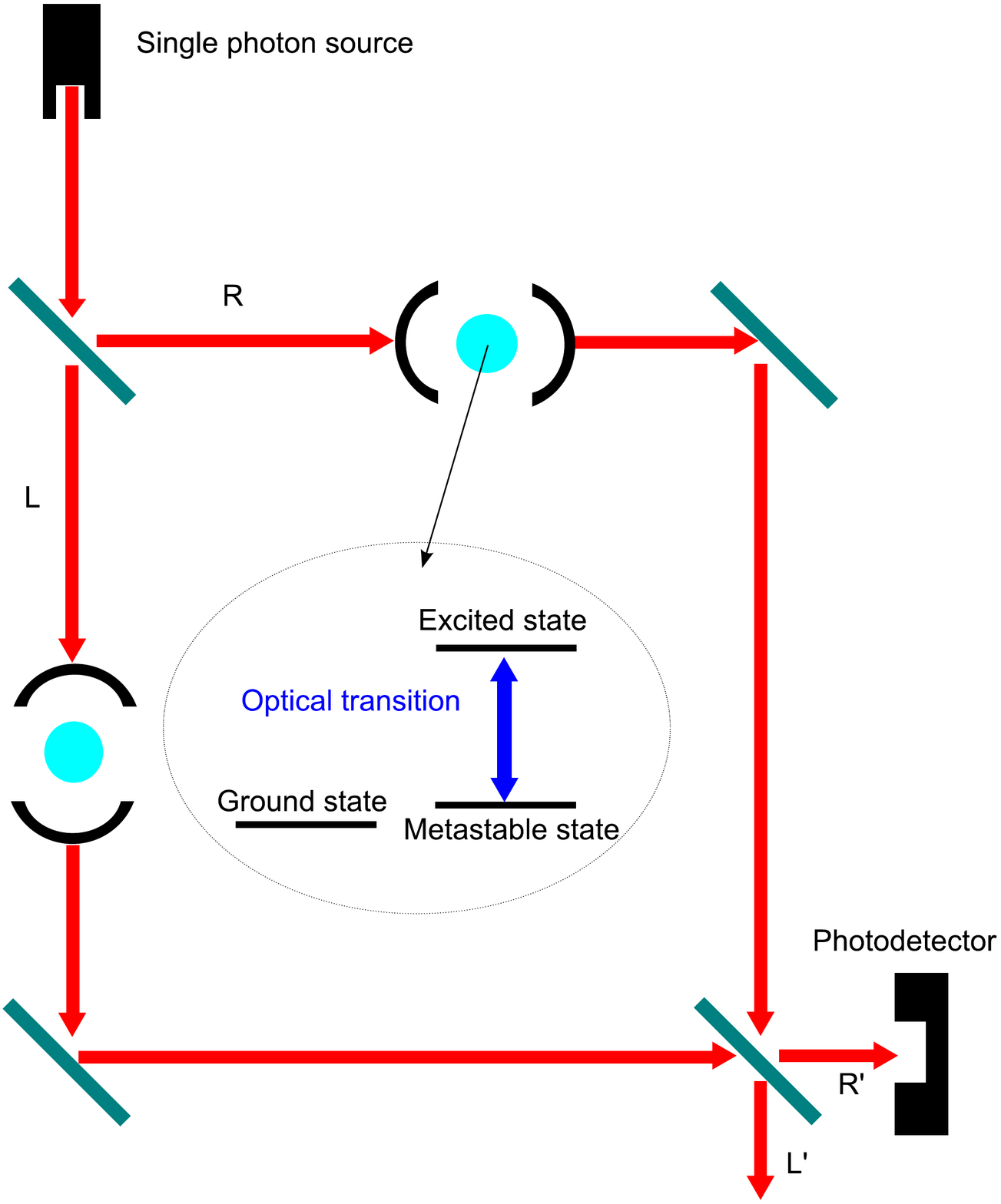}
   \end{center}
   \caption{Schematic of an apparatus for the EO. A
     half mirror splits a single photon (or a weak coherent laser ) into two paths. A
     matter qubit (denoted `atom', but maybe quantum dot etc) is confined in a cavity on each path. Each atom has an L-type structure and has three
   quantum states
   $|0\rangle $, $|1\rangle $, and $|e\rangle $ as shown in this
     schematic. The photon frequency matches a cavity mode and
     so the photon transmission probability at the
     cavity can approach unity.
     Ultimately the photon goes through another splitter and, depending on output port, may be measured by a
     photon detector. A detection event projects the atoms into an entangled
     state. 
     Since there is a large detuning between a cavity
     frequency and the atomic transition energy, the optical transition
     is
     suppressed, and therefore the excited state can have a large energy fluctuation without decreasing the fidelity of the entanglement.
     }
    \label{2bloch}
  \end{figure}

We begin by describing a simplified scenario and then consider realistic imperfections. We will use the term `atom' to refer to a generic optically active qubit, which may in fact be a quantum dot or crystal defect (we discuss such possibilities presently). 
 In essence, a single photon detuned from the
 atomic transition induces a relative phase between
 the atomic states. 
 Since the atom(s) never undergo a transition to the optically excited state, the scheme
 can be extremely robust against fluctuations in such states.
 Surprisingly, even when the environmental coupling strength is the same order
 of magnitude as the atom-photon coupling strength, one can still
 generate a high-fidelity entanglement with a reasonable
 success probability.
 
 We assume that a matter qubit in a cavity has an L-type structure and has three
 quantum states
 $|0\rangle $, $|1\rangle $, and $|e\rangle $. A state $|1\rangle $
 is optically active and coupled to a noisy excited state $|e\rangle $
 whereas the state $|0\rangle $ is not optically active (see Fig.~\ref{2bloch}).
 Two such atoms are remotely located, each within a cavity (we discuss the case without cavities presently). Initially the atoms  are prepared as $|+\rangle _{L}|+\rangle
 _R=\frac{1}{2}(|0\rangle _L|0\rangle _R+|0\rangle _L|1\rangle
 _R+|1\rangle _L|0\rangle _R+|1\rangle _L|1\rangle _R)$ where $L$ and
 $R$ denote the location of each cavity.
  When a frequency of a cavity mode is detuned from an atomic transition,
  a simplistic effective interaction Hamiltonian can be written as~\cite{holland1991quantum}
    \begin{eqnarray}
   H_{\text{eff}}\simeq 
    \frac{g^2}{\Delta }(|e\rangle \langle e|-|1\rangle \langle 1|) \hat{a}^{\dagger }\hat{a}\label{effective}
  \end{eqnarray}
   where $g$ is the coupling strength of the
 cavity and $\Delta $ is the detuning between the cavity mode and the atomic
 transition. This effective Hamiltonian can be derived from the standard
 Jaynes-Cummings Hamiltonian in the limit of large detuning.

We suppose that a single photon is split by a half mirror into two paths, along which the two cavities lie symmetrically. 
Importantly the photons are of a frequency that is significantly detuned from the atomic transitions; the cavity mode frequencies are matched to the frequency of the photons. Despite the detuning, when a photon interacts with an atom in the optically active atomic state $|1\rangle $ a certain phase is acquired. However there is no interaction between the photon and the atom for an
 optically inactive state $|0\rangle $.
 
 Subsequent to the atom-photon interaction, our state $|+\rangle _{L}|+\rangle_R$ has evolved to 

\begin{eqnarray}
\frac{1}{2\sqrt{2}}\Big{(}
 |0\rangle _L|0\rangle _R
 (\hat{a}^{\dagger}_L+\hat{a}^{\dagger}_R)+|0\rangle _L|1\rangle _R
 (\hat{a}^{\dagger}_L+\hat{a}^{\dagger}_Re^{i\theta })
 + \nonumber \\ 
 |1\rangle
 _L|0\rangle _R (\hat{a}^{\dagger}_Le^{i\theta } +\hat{a}^{\dagger}_R)
 +|1\rangle _L|1\rangle _R (\hat{a}^{\dagger}_Le^{i\theta }
 +\hat{a}^{\dagger}_Re^{i\theta })\Big{)}
 \end{eqnarray}
 where $\hat{a}$
 ($\hat{a}^{\dagger}$) denote an annihilation (creation) operator of the
 photon.
 The phase is effectively described as $\theta \simeq \frac{g^2}{\Delta
 } t$
where $t$ is an interaction time.
 Since the frequency of the photon is assumed to be centered
 around the cavity frequency in a narrow range, the photon can be transmitted through the
 cavity without reflection~\cite{collett1984squeezing}.
 Finally the photon goes through another half mirror to change the mode
 $\hat{a}^{\dagger }_L$($\hat{a}^{\dagger }_R$) into
 $\hat{a}^{\dagger }_{L'}=\frac{\hat{a}^{\dagger }_{L} +\hat{a}^{\dagger
 }_{R}}{\sqrt{2}}~$($\hat{a}^{\dagger }_{R'}=\frac{\hat{a}^{\dagger }_{L} -\hat{a}^{\dagger
 }_{R}}{\sqrt{2}}$) where $L'$ and $R'$ denote the output ports of the
 splitter, and we obtain 
 \begin{eqnarray}
 \frac{e^{i\theta }}{2}|0\rangle _L|0\rangle _R \hat{a}^{\dagger }_{L'}
 +|0\rangle _L|1\rangle _R\frac{e^{\frac{1}{2}i\theta }}{2}(\cos \frac{\theta }{2}  \hat{a}^{\dagger
 }_{L'}+i\sin \frac{\theta }{2}  \hat{a}^{\dagger }_{R'})
 + \nonumber \\
 \frac{e^{\frac{1}{2}i\theta }}{2}
 |1\rangle _{L}|0\rangle _{R} (\cos \frac{\theta }{2}  \hat{a}^{\dagger
 }_{L'}-i\sin \frac{\theta }{2}  \hat{a}^{\dagger }_{R'})
 +|1\rangle _L|1\rangle _R   \hat{a}^{\dagger }_{L'}.
 \end{eqnarray}

 If the atoms were not present, or were in either the definite state $|0\rangle _{L} |0\rangle _{L}$, or the state $|1\rangle _{L} |1\rangle _{L}$, then the photon would certainly exit from the left port of the second splitter. However, because of the internal phase shifts, the photon may exit from the right port -- this represents a successful entanglement. The success probability is $\frac{1}{2}\sin ^2\frac{\theta }{2}$ which
 provides us with a maximum value $0.5$ for $\theta =\pi $, and regardless of $\theta$
 this measurement projects the atomic state into an entangled state
 $\frac{1}{\sqrt{2}}|0\rangle _L|1\rangle _R
 -\frac{1}{\sqrt{2}}|1\rangle _L|0\rangle _R$. In this example the
 initial state was assumed to be $|+\rangle _L|+\rangle _R$, however one can perform this
 operation on arbitrary initial states, and success results in a parity
 projection i.e. a projector onto a odd parity two-qubit subspace,
 which is one of the typical EOs for distributed quantum computation~\cite{BK01a,BPH01a,Benjamin:2006p358,PRAmatsuzaki2010distributed}.
 
In practice no single photon source will be ideal. Therefore we now consider both an imperfect single photon source, and a weak coherent laser, as alternatives. The ideal single photon source should emit one and only one
photon when the device is triggered, which can be realized to some approximation by exploiting the photon
antibunching effect~\cite{KDM01a}.
With current technology
it is inevitable that the pulse
generated by a source
may contain either no photons, or multiple photons, with finite
probability.
Suppose that $P_m$ denotes the probability of emitting $m$ photons. Importantly, finite $P_0$
will not give rise to errors; in effect it adds to the probability of
photon loss and will be registered as a failure by the
detectors. However, finite $P_m$ with $m\geq 2$ will give rise to errors; for example,
for $m=2$ a likely occurrence is that one photon will be lost and the other
registered by the detector, in which case we would wrongly conclude that
the normal EO has succeeded.
Taking the worst case that such emissions can make
the state orthogonal to the target state, the fidelity is bounded as
$F\geq 1-\sum_{m\geq 2} P_m$. Fortunately,
it is possible to make the probability of such events rather small. For example, 
Ref.~\cite{LM01a} reports a single photon source whose $P_0$ and 
$P_2$ are $14\%$ and $0.08\%$ respectively (with negligible chance of higher $m$).

For near future demonstrations of the protocol described here, it may be appropriate to utilize an even less ideal source: namely a weak coherent laser.
 A coherent state is described as $|\alpha \rangle
 =e^{-\frac{1}{2}|\alpha |^2}\sum_{n=0}^{\infty
 }\frac{1}{\sqrt{n!}}\alpha ^n|n\rangle $ where $|\alpha |^2 $ denotes
 a mean number of the photons and $|n\rangle $ denote a number state of
 the photon.
 A coherent state $|\alpha \rangle $ acquires a phase as $|\alpha
 e^{i\theta }\rangle $ when the atomic state is optically active.
 So, taking the same operation as a single photon mentioned above, the state following the final beam-splitter will be 
 $ \frac{1}{2}|\alpha \rangle _{L'}|\text{vac}\rangle _{R'}|0\rangle _L|0\rangle _R
  +\frac{1}{2}|\alpha
  e^{\frac{i\theta}{2} }\cos \frac{\theta }{2}
  \rangle _{L'} |\alpha
  e^{\frac{i\theta}{2} }\sin \frac{\theta }{2}
  \rangle _{R'} |0\rangle _L|1\rangle _R
 +\frac{1}{2}|\alpha
  e^{\frac{i\theta}{2} }\cos \frac{\theta }{2}
  \rangle _{L'} |-\alpha
  e^{\frac{i\theta}{2} }\sin \frac{\theta }{2} 
  \rangle _{R'} |1\rangle _L|0\rangle _R
  +\frac{1}{2}|\alpha \rangle _{L'}|\text{vac}\rangle _{R'}|1\rangle
 _L|1\rangle _R  $ where $|\text{vac}\rangle _{R'}$ denotes a vacuum
 state of mode $R'$.
Detecting a single photon
of the mode $R'$ projects the atomic
state into an entangled state $|\psi ^{(-)}\rangle =\frac{1}{\sqrt{2}}|0\rangle _L|1\rangle _R
-\frac{1}{\sqrt{2}}|1\rangle _L|0\rangle _R$, while detecting two
 photons project the state into $|\psi ^{(+)}\rangle =\frac{1}{\sqrt{2}}|0\rangle _L|1\rangle _R
+\frac{1}{\sqrt{2}}|1\rangle _L|0\rangle _R$.
 So a photon number resolution device can project the atomic state into
 a Bell state. However, if one wished to operate the protocol without the availability of a reliable
  photon number resolution device, one could still obtain entanglement between the
matter qubits. 
In that case when the detector at the output $R'$ registers (a non-zero but unknown number of) photons, the state of
the matter qubit becomes a classical mixture of the target state $|\psi ^{(-)}\rangle$
and the other error states $|\psi ^{(+)}\rangle$.
A fidelity $F$ and a success probability $P$
are scaled as
$F= 1-\frac{1}{2}|\alpha |^2 \sin ^2\frac{\theta }{2}+O(|\alpha |^4)$
and $P= \frac{1}{2}|\alpha |^2 \sin ^2\frac{\theta }{2}+O(|\alpha |^4)$ for
$|\alpha |^2 \ll 1$, respectively. 
Here, at the expense of
success probability, a weaker coherent state can increase the fidelity
by guaranteeing that it is unlikely for multiple photons to reach the detector.
So
one can obtain high fidelity entanglement by using a weak coherent
state.
Although the success probability becomes relatively low
due to
the trade-off relationship between the fidelity and the success
probability,
a coherent
laser is much easier to construct than a single photon source, and
so this scheme using a coherent state should be feasible even with current
technology.

Photon loss is a major source of error in the most of
previously proposed EOs~\cite{CEHM01a,LBK01a,BPH01a,BKPV01a,Fengetal01a,laddetal2006hybrid},
 including the ingenious schemes \cite{laddetal2006hybrid,azuma-optimal,azuma-coherent}
which, like the present scheme, operate by inducing a phase. In effect we propose to exploit the fact that single photon sources are becoming a mature technology and can therefore be substituted for the classical source in that previous scheme, with the benefit that photon loss will now be detected and thus prevented from impairing fidelity.
Also, it is worth mentioning that cavities are not essential for our
scheme as long as one can achieve a strong coupling between a photon and
an atom. For example, recently, strong interaction between light and a
single atom in a free space has been demonstrated by using a lens~\cite{teyetal2008strong}, and a
phase shift of a weak coherent state about $1 ^\circ$, which is induced
by a single atom, has
been observed experimentally~\cite{aljunidetal2009phase}.
Since our analysis above can be directly applied to
the case of free space, these
experiments also demonstrate the feasibility of performing our scheme without the
need for a cavity.

We now present a more detailed analysis. In the previous description we adopted an approximation that there is no optical transition
because of a large detuning between the atomic transition and the
frequency of the photon.
However, even when the detuning $\Delta $ is large, there is a non-zero
 probability for the
 state to be excited, which might affect the fidelity of the EO.
 To include this effect,
 we use the following Hamiltonian instead of the effective Hamiltonian (\ref{effective})
 \begin{eqnarray}
  &&H=\sum_{j=L,R}\big{(}\frac{\omega}{2} \hat{\sigma }^{(j)}_z+ \nu
   \hat{a}_j^{\dagger }\hat{a}_j\big{)}\nonumber \\
  &+&\sum_{j=L,R}g\big{(}\hat{\sigma }^{(j)}_+ \hat{a}_j+ \hat{\sigma
   }^{(j)}_-\hat{a}^{\dagger }_j\big{)}
   -i\Gamma \big{(}\hat{a}^{\dagger }_{L}\hat{a}_{L} +\hat{a}^{\dagger
   }_{R}\hat{a}_{R}\big{)}\ \ \ \ \ \
 \end{eqnarray}
 where $\omega $, $\nu $, $g$, and $\Gamma $ denote the atomic transition
 energy, the cavity frequency, the coupling strength of the cavity, and the
 decay rate of the
 cavity respectively.
   \begin{figure}[h]
   \begin{center}
    \includegraphics[width=6.5cm]{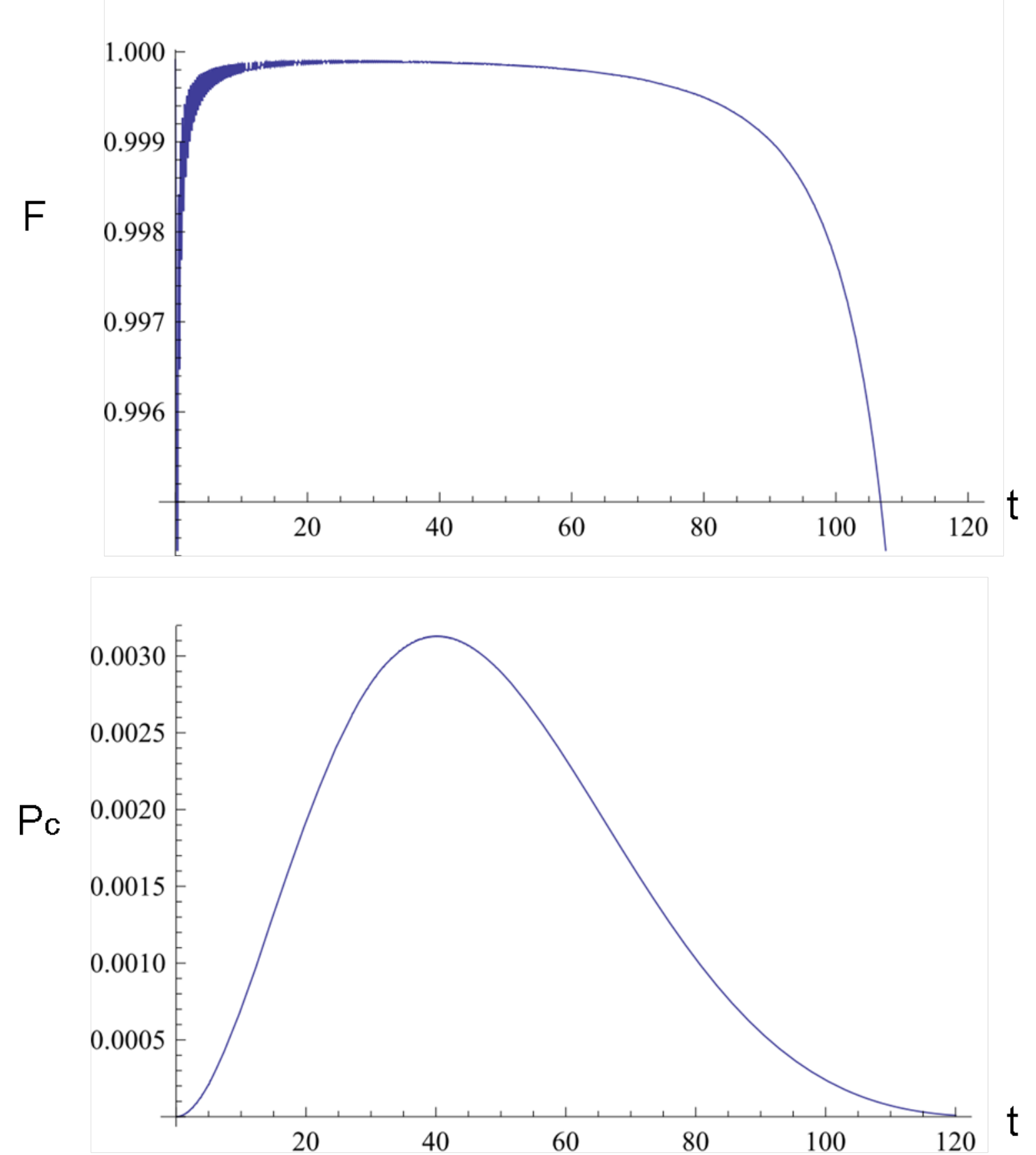}
   \end{center}
    \caption{ Upper: The fidelity of an EO using a
    single photon source is plotted, where $t$ denotes the time of the detection by the
    photon detector. Lower: $P_c(t)$, the probability of the desired detector `click' per unit time. 
    The average fidelity is calculated as
    0.998 and the total success probability is
    0.178.
    We set parameters as $\Delta =20g$, $\Gamma =\frac{1}{\pi
     }\frac{g^2}{\Delta }$, $\lambda =0.1g$, and $g=1$ where $\Delta $,
    $\Gamma $, $\lambda $, and $g$ denote a detuning, a cavity decay rate, an environmental
    coupling, and the atom-cavity coupling respectively.}\label{single-fidelity}
   \end{figure}
 Note that we have added not only the standard
 Jaynes-Cummings Hamiltonian (the first and second term) but also a
 decay term (the last term) to describe the conditional dynamics when no photon is
 measured at the detector~\cite{carmichael1993open}. Importantly,
 $\frac{1}{\Gamma}$ gives the characteristic time during which the photon in the
 cavity interacts with the atom.
 For a large detuning, the effective Hamiltonian
 (\ref{effective}) is a good approximation, and therefore it is necessary to satisfy 
 $\frac{g^2}{\Delta
 }\frac{1}{\Gamma}\simeq \pi $ so that one can obtain
 a reasonable success probability.
  To satisfy this requirement,
a strong
 coupling regime $g\gg \Gamma $ is in turn required since we will employ a 
  $\Delta $ which is much larger than the coupling strength $g$
(to prevent the state from being excited).

 To model the effect of the energy fluctuation of the excited state,
 we use a Lindblad master equation as follows:

 \begin{eqnarray}
  \frac{d\rho }{dt}=-i(H\rho -\rho  H^{\dagger })
   -\lambda \sum_{j=L,R}[\ |e\rangle _j\langle
   e|,\ [\ |e\rangle _j\langle e|\ ,\rho ]].
 \end{eqnarray}
The solution of this master equation provides us with a density matrix
of the state while the detector registers no photons. When the detector registers an event in mode $R'$ observed at time $t$, the state is projected onto $\rho
   _{\text{final}}(t)=\hat{a}_{R'}\rho(t)
\hat{a}^{\dagger }_{R'}$ discontinuously.
For $\Delta =20g$, $\Gamma =\frac{1}{\pi
   }\frac{g^2}{\Delta }$, and $\lambda =0.1g$, we have plotted fidelity
   $F(t)=\langle \psi (t)_{\text{bell}}|\rho _{\text{final}}|\psi
   _{\text{bell}}\rangle $
   in Fig. \ref{single-fidelity}
   where $|\psi _{\text{bell}}\rangle
   =\frac{1}{\sqrt{2}}(|01\rangle -|10\rangle )$.
   This figure shows that, after the fidelity takes a maximum value,
   it decays due to the energy fluctuation.
   By taking an
   average of the fidelity, we obtain
   $F_{\text{average}}=\int_{0}^{\infty
   }P_c(t)F(t)dt\simeq 0.998$ where $P_c(t)$ denotes a
   probability of clicking the photon at a time $t$ (also shown in the Figure).

         \begin{figure}[h]
    \begin{center}
     \includegraphics[width=6.5cm]{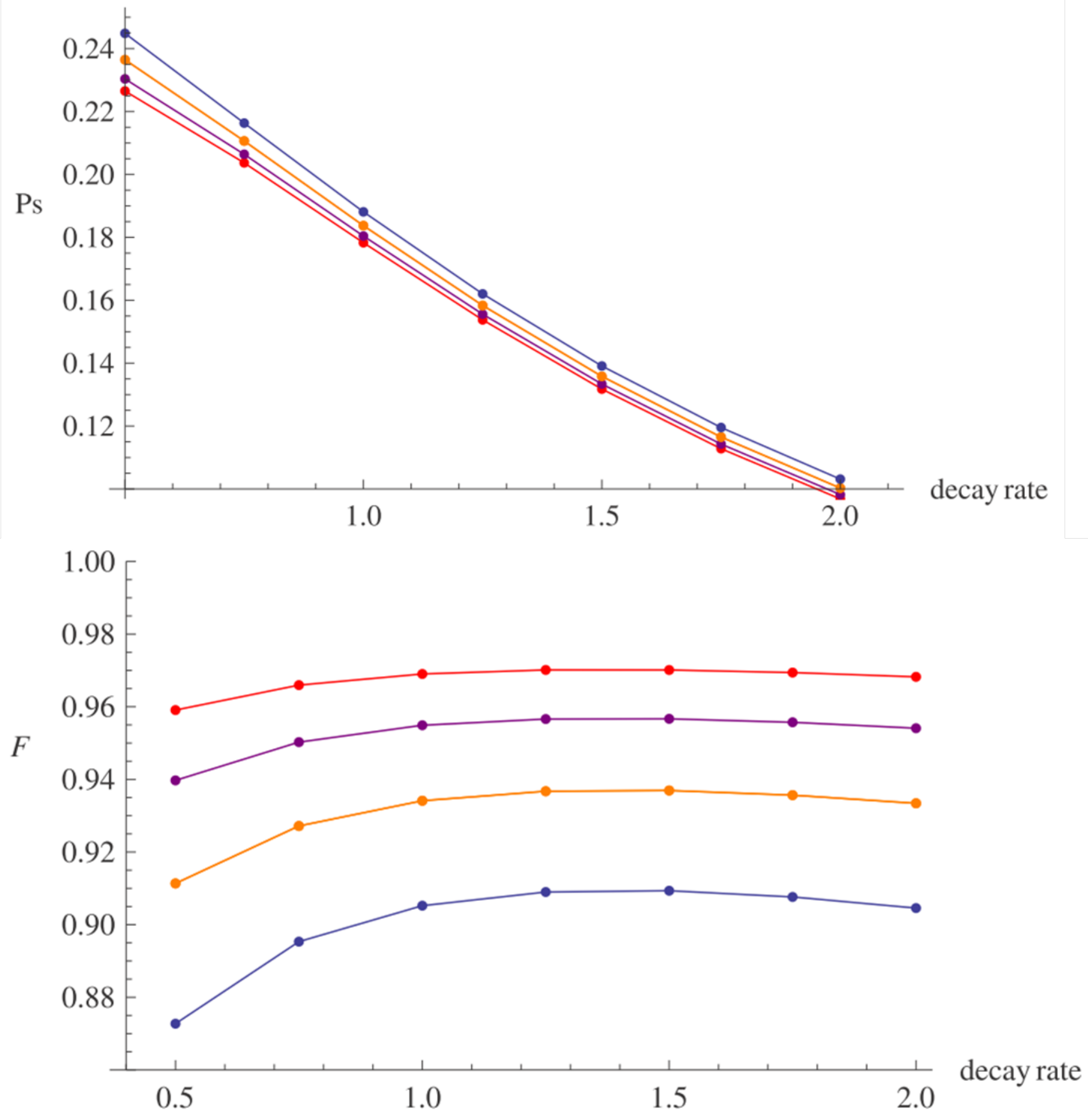}
    \end{center}
         \caption{ The success probability and fidelity of an EO using a
    single photon source is plotted against the normalized decay rate of a
       cavity with $\gamma = \Gamma /(\frac{1}{\pi}\frac{g^2}{\Delta })$. 
       The highest line (blue) in the upper graph and the lowest line
       (blue) in the below graph
       are the case of $\Delta =15g$, and  the other lines are the case of
       $\Delta =20g, 25g, 30g$ (orange, purple, red), respectively.  Also, we have fixed the environmental
       coupling and the atom-cavity coupling as $\lambda =g$.}\label{scale-single}
    \end{figure}
  By integrating the success probability over the
   time $t$, we obtain a total success probability as $0.177$ and
   this success probability is large enough to grow a resource such as a cluster state
  on a practical time
   scale~\cite{BK01a,DR01a,GKE02a,UTprl} without employing a brokering strategy\cite{Benjamin:2006p358}.
Furthermore, we have studied how the decay rate changes the total success
   probability and the average fidelity.
   In Fig. \ref{scale-single}, we have plotted the
   average fidelity and success probability against a normalized decay
   rate $\gamma = \Gamma /(\frac{1}{\pi}\frac{g^2}{\Delta })$ when the
   environmental coupling strength $\lambda $ is the same value as the coupling
   strength $g$.
   The fidelity takes a maximum value when the normalized decay rate is
around $\frac{3}{2}$.
   Even when the coupling strength
   of the noise is the same order of the magnitude as
 the atom-cavity coupling strength, our scheme can still
 generate entanglement with a reasonable
 success probability and good fidelity.
     \begin{figure}[h]
    \begin{center}
     \includegraphics[width=6.5cm]{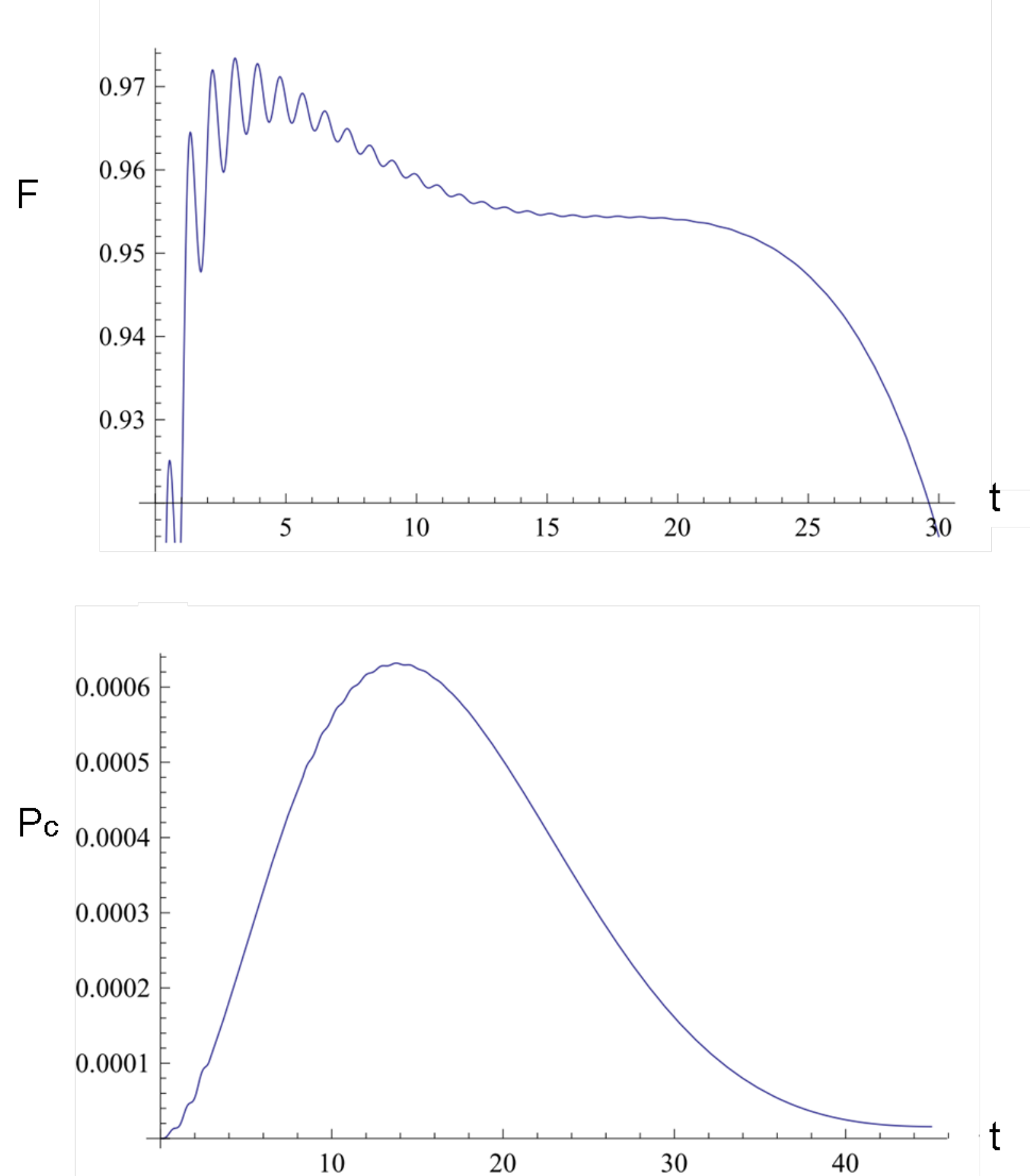}
    \end{center}
      \caption{ Upper: The fidelity of an EO using a
    weak coherent state is plotted, where $t$ denotes the time of the detection by the
    photon detector. Lower: $P_c(t)$, the probability of the desired detector `click' per unit time. 
  The total success probability is
      $0.0128$ and the average fidelity is $0.939$.
    We set parameters as $\Delta =7g$, $\Gamma =\frac{1}{\pi }\frac{g^2}{\Delta }$,
      $\lambda =0.5 g$, $\alpha =0.2$, and $g=1$ where $\Delta $,
    $\Gamma $, $\lambda $, $\alpha $, and $g$ denote detuning, decay rate of the cavity, environmental
    coupling, amplitude of the coherent state, and the atom-cavity coupling respectively.
        }\label{coherent}
    \end{figure}

 In Fig.~\ref{coherent}, we show the case of using a weak coherent state.
 For $\Delta =7g$, $\Gamma =\frac{1}{\pi }\frac{g^2}{\Delta }$,
   $\lambda =0.5 g$, and $\alpha =0.2$, 
 the average fidelity is $0.939$ and the total success probability is
 $0.0128$. While this success probability may be too low to support
 universal quantum computing (without the use of additional memory
 qubits \cite{Benjamin:2006p358}),
 it is however certainly high enough to permit smaller scale applications or a comprehensive experimental demonstration of the protocol. 
 

 Finally, we describe possible experimental realizations.
 A quantum dot (QD) defined on n-type GaAs heterostructures
 is one of the candidates.
 A polarized photon can selectively drive
  one of the electron spin states into an excited state called a trion, and so it is
 possible to construct the needed L-type structure.
 Moreover, strong coupling with a photon has already been
 realized in a QD in a semiconductor microcavity
 where
 $g=80\mu $ev and $\Gamma =33 \mu$ev~\cite{reithmaieretal2004strong,reitzensteinetal2007alas}.
 Another candidate is a p-type GaAs QD
 where a single hole has two spin states.
 Importantly, a long spin
 relaxation time of the spin $T_1\simeq 1$ ms has been demonstrated~\cite{gerardotetal2008optical}, and
 the same order of spin dephasing time $T_2$ is theoretically
 predicted in the hole spin states~\cite{golovach2004phonon}, which is
 much longer than the electron spin dephasing time $T_2\simeq 10$ ns
 in a n-type QD~\cite{pettaetal2005coherent}.
 These otherwise attractive systems are impaired by the optical emission of the hole spin
 states which has a large line width implying the excited states are more noisy
 than
 n-doped QDs~\cite{gerardotetal2008optical}. This is therefore a very relevant class of system for the present scheme, by which
  it is possible to perform high fidelity EOs despite the noisy excited state.

Nitrogen vacancy (NV) centers
in diamond are a second system with promising properties (including an electron dephasing time 
about a millisecond at a room temperature~\cite{gaebeletal2006room}), marred by an optically excited state with strong phonon interactions.
Therefore this system is also very relevant to the present scheme, although a suitable strong coupling with a photon through a cavity has not yet been demonstrated.

 In conclusion, we have described a scheme to entangle distant matter qubits even when those qubits suffer severe energy fluctuations.  
 Many optically active solid state systems suffer
unstable excited states, and our scheme provides a
 practical way to overcome such typical issues. The authors thank J.M. Smith for a useful discussion. This research is supported by the National Research Foundation and Ministry of Education, Singapore.
 JF acknowledges support from Merton College.
YM is supported by the Japanese Ministry of Education, Culture, Sports, Science and 
Technology.

\end{document}